%% file: main.tex
\documentclass[journal]{IEEEtran}
\input{macros.tex}

\title{Robust, Informative Human-in-the-Loop Predictions via Empirical Reachable Sets}
\author{Katherine Driggs-Campbell,~\IEEEmembership{Student Member,~IEEE,} Roy Dong,~\IEEEmembership{Student Member,~IEEE,} \\ S. Shankar Sastry,~\IEEEmembership{Member,~IEEE,}      and~Ruzena~Bajcsy,~\IEEEmembership{Member,~IEEE}% <-this % stops a space
\thanks{K. Driggs-Campbell, R. Dong, S.S. Sastry, and R. Bajcsy are with the Department of Electrical Engineering and Computer Sciences, University of California, Berkeley, Berkeley, CA, 94720 USA e-mail: \{krdc,roydong,sastry,bajcsy\}@eecs.berkeley.edu.}% <-this % stops a space
\thanks{Manuscript received \today.}}

\begin{document}

\maketitle

\begin{abstract}
In order to develop provably safe human-in-the-loop systems, accurate and precise models of human behavior must be developed.  
In the case of intelligent vehicles, one can imagine the need for predicting driver behavior to develop minimally invasive active safety systems or to safely interact with other vehicles on the road.
We present a optimization based method for approximating the stochastic reachable set for human-in-the-loop systems.
This method identifies the most precise subset of states that a human driven vehicle may enter, given some dataset of observed trajectories.
We phrase this problem as a mixed integer linear program, which can be solved using branch and bound methods.
The resulting model uncovers the most representative subset that encapsulates the likely trajectories, up to some probability threshold, by optimally rejects outliers in the dataset.
This tool provides set predictions consisting of trajectories observed from the nonlinear dynamics and behaviors of the human driven car, and can account for modes of behavior, like the driver state or intent.
This allows us to predict driving behavior over long time horizons with high accuracy.
By using this realistic data and flexible algorithm, a precise and accurate driver model can be developed to capture likely behaviors.
The resulting prediction can be tailored to an individual for use in semi-autonomous frameworks or generally applied for autonomous planning in interactive maneuvers.
\end{abstract}

%\begin{IEEEkeywords}
%IEEE, IEEEtran, journal, \LaTeX, paper, template.
%\end{IEEEkeywords}

% \fi
% \IEEEpeerreviewmaketitle

\input{sections/introduction.tex}

\input{sections/modeling_methods.tex}

\input{sections/algorithm_evalution.tex}
\input{sections/application.tex}
\input{sections/discussion.tex}

%\section*{Acknowledgment}
%\todonote{The authors would like to thank...}

\clearpage
\newpage

\bibliographystyle{IEEEtran}
\bibliography{krdcBibFile,vijayBibFile}

%
%\begin{IEEEbiography}{Katherine Driggs-Campbell}
%\end{IEEEbiography}
%
%\begin{IEEEbiography}{Roy Dong}
%\end{IEEEbiography}
%
%\begin{IEEEbiography}{S. Shankar Sastry}
%\end{IEEEbiography}
%
%% insert where needed to balance the two columns on the last page with
%%\newpage
%
%\begin{IEEEbiography}{Ruzena Bajcsy}
%\end{IEEEbiography}

\end{document}

%% file: macros.tex
\usepackage{graphics} % for pdf, bitmapped graphics files
\usepackage{amsmath} % assumes amsmath package installed
\usepackage{amssymb}  % assumes amsmath package installed
\usepackage{multirow}% http://ctan.org/pkg/multirow
\usepackage{hhline}
\usepackage{algorithm}
\usepackage[noend]{algpseudocode}
\makeatletter
\def\BState{\State\hskip-\ALG@thistlm}
\makeatother
\PassOptionsToPackage{hyphens}{url}
\usepackage{dsfont}
\usepackage{algorithmicx}
\usepackage{algpseudocode}
\usepackage{multirow}
\usepackage{epigraph}
\usepackage{booktabs}
\usepackage{scrextend}
\usepackage{rotating}

\usepackage[utf8]{inputenc}
\usepackage{import}
\usepackage{amsfonts}
\usepackage{amsmath}
\usepackage{csquotes}

\usepackage{wrapfig}
\usepackage{graphicx}
\usepackage{caption}
\usepackage{subcaption}

\usepackage{array}
\usepackage{graphics} % for pdf, bitmapped graphics files
\usepackage{url}
\usepackage{dsfont}
\usepackage{algorithmicx}
\usepackage{algpseudocode}
\usepackage{graphicx,caption}
\usepackage{amsmath,amssymb,stmaryrd,mathtools}
\usepackage{flushend}
\usepackage{algorithm,algorithmicx}
\usepackage{acronym}
\usepackage{color,xcolor}
\graphicspath{{./figures/}}

\newcommand{\ignore}[1]{}
\DeclareFontFamily{U}{MnSymbolC}{}
\DeclareSymbolFont{MnSyC}{U}{MnSymbolC}{m}{n}
\DeclareFontShape{U}{MnSymbolC}{m}{n}{
	<-6>  MnSymbolC5
	<6-7>  MnSymbolC6
	<7-8>  MnSymbolC7
	<8-9>  MnSymbolC8
	<9-10> MnSymbolC9
	<10-12> MnSymbolC10
	<12->   MnSymbolC12%
}{}
\DeclareMathSymbol{\powerset}{\mathord}{MnSyC}{180}

\newcommand{\lbar}[1] {\underbar{$#1$}}
\newcommand{\ubar}[1] {\overline{#1}}

\newcommand{\AM}[1]{\underset{#1}{\text{argmin}}\;}

%% file: sections/introduction.tex
\section{Introduction}
\label{sec:intro}

When considering human-robot interaction and human-in-the-loop systems, one of the primary concerns is how to estimate and guarantee safety.  From a design perspective, there are many different approaches that can be used.   Some robotic systems approach safety from a mechanical point of view by creating systems that physically cannot harm the human \cite{baxter2014}.  Another approach is to develop controllers and sensor systems that can guarantee safety for a given system \cite{Gillula2012}.  
However, when considering systems that involve or interact with humans (e.g. human-in-the-loop systems), deriving safety boundaries and assessments is not a simple task, as many of the classical assumptions break down when giving the human influence in the system.  This is due to the fact that human actions and behaviors are often unpredictable and cannot easily be described by known distributions or by normal dynamical methods \cite{driggscampbell2015}.  
Another difficulty comes from the computational complexity that arises from humans possible action spaces.  To compensate for this, simplified models are used to represent the system without proper metrics to measure how well the model matches real-world behavior. 

\begin{figure*}
	\centering
	\includegraphics[width=.8\textwidth]{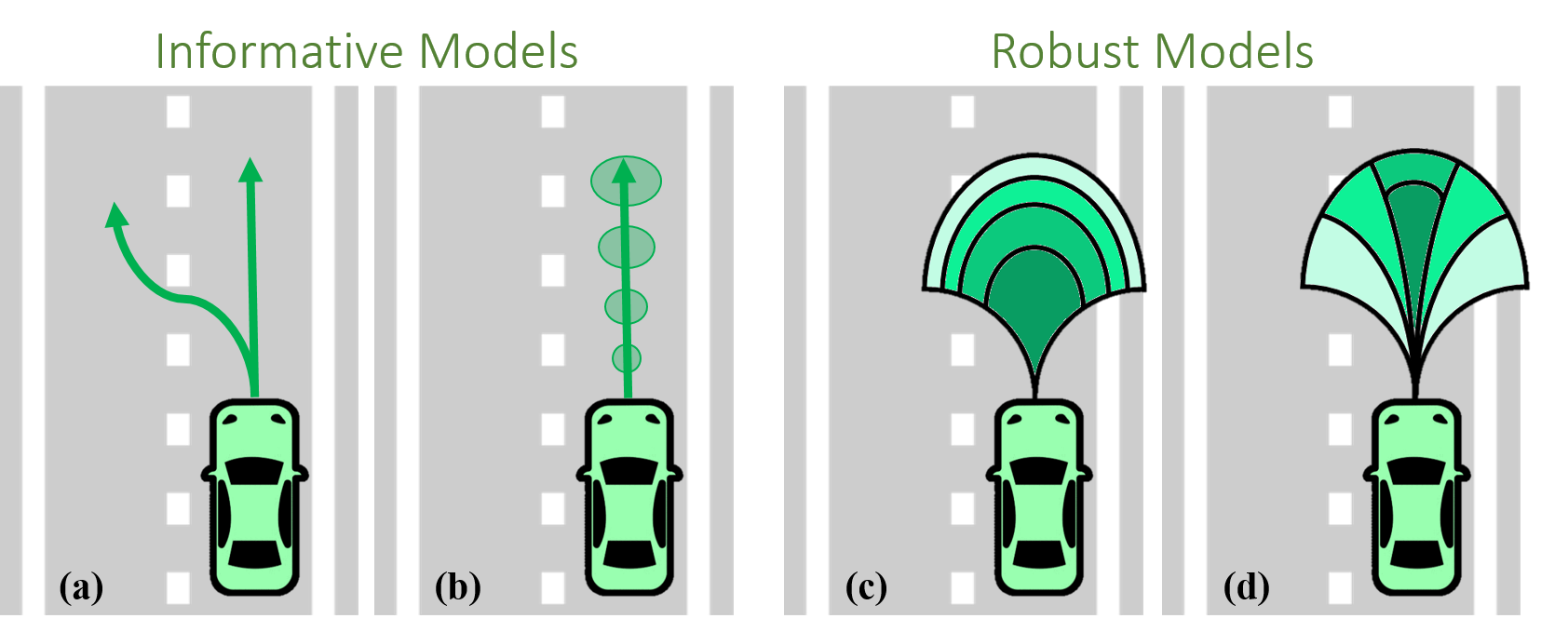}
	\caption{\textbf{Informative versus Robust Modeling.} (a) Maximally informative, but least robust prediction. (b) Informative prediction with assumptions over distributions on human behavior. (c) Maximally robust prediction, requiring exact model and disturbance bounds and tends to be over-conservative. (d) Visualization of prediction that identifies useful subsets of reachable set, balancing robustness with informativeness.}
	\label{fig:informativeVSrobust}
\end{figure*}

To develop provably safe human-in-the-loop systems, first an informative and accurate model of the human must be developed that can be incorporated into control frameworks.  
In deriving the modeling methodology, we consider two possible approaches to modeling human behavior: informative and robust methods, as visualized in Figure \ref{fig:informativeVSrobust}.

\subsection{Informative Predictions}
In order to have safe and interactive systems, predictive modeling is incredibly important \cite{driggscampbell2017}.
Ideally, for each obstacle in the environment, the exact future trajectory would be able to be uncovered for all scenarios.
Having a precise trajectory would maximize the \emph{informativeness} or the \emph{utility} of the prediction.

However, given the randomness of human motion, it is unlikely that the precise trajectory will be uncovered uniquely \cite{vasudevan2012}.
While this has been applied to very specific situations under strict assumptions \cite{anderson2010}, the probability of this functioning reliably is negligible.
In the realm of intelligent vehicles, many works have developed models attempt to predict the exact trajectory, but either do not generalize well or cover unknown situations \cite{houenou2013vehicle,werling2012optimal}.

To gain more utility, probabilistic approaches have been applied to allow some uncertainty about a nominal trajectory \cite{gindele2010probabilistic,Wiest2012}.
Again, this requires many assumptions on the distribution over driving behavior, which is often violated \cite{krdc2013}.
Stochastic models have also been developed, but make many assumptions on the underlying model of human behavior (e.g. Markov Decision Processes assume humans satisfy the Markov property \cite{abbeel2004}) or on the distribution on human actions \cite{driggscampbell2015}.

\subsection{Robust Predictions}
In contrast to informative models, reachable sets maximize the \emph{robustness} and \emph{accuracy} of the prediction.
This modeling methodology is inspired by the control theoretic tool of forward reachability, which gives lovely provable guarantees of encapsulating the systems behaviors, assuming the model and disturbances are well known.
Given these assumptions, these and related methods can provide certificates that give an exact proof of safety \cite{prajna2007framework}.
As a consequence, these methods tend to be over-conservative, meaning that the prediction is highly accurate, but not informative.

In order to utilize these techniques, many assumptions must be made on the model being used.
There has been a great deal of work aiming to a address these issues by considering stochastic reachability or by applying safe learning techniques. 

For human controlled systems, the disturbances are often difficult to model and use in control frameworks \cite{shia2013}.
The disturbances, however, are crucial in robust modeling--if the assumed disturbance bounds do not globally capture the true disturbance, reachability methods can no longer guarantee safety.  
On the other hand, if the disturbances are over approximated, the resulting control will be over-conservative \cite{driggscampbell2017}.

To address this issue, there has been growing interest in learning these disturbances online to reduce the conservativeness of these methods \cite{gillula2013}.
In \cite{akametalu2014}, the authors designed a safe online learning framework to both learn disturbances and modeling errors, while applying reinforcement learning for control.

A key inspiration for considering the reachability framework is the Volvo City Safety system, a successful semiautonomous system that relies on such reachable sets to mitigate collisions.  When driving in the city (below 35 miles per hour), the system calculates the forward reachable set of the vehicle for the future 500ms and anticipates collisions by checking to see if a detected object is within that set \cite{coelingh2007collision}.   As noted, this method does not work at high speeds as the reachable set of the vehicle itself becomes too large, leading to an overly invasive system.  
When considering high speeds, the human can no longer be considered as a disturbance in the system, as the driver has significant influence over the future trajectories of the vehicle.
Ideally, the system would function at high speeds and consider the \emph{likely} actions of the human by modeling the driver to create a more informative reachable set.

\subsection{Our Contributions}
Here, we present a method for identifying the subset of the reachable set that is useful up to some probability threshold, which we will call the \emph{empirical reachable set}.
The algorithm estimates the nonparametric distribution empirically induced by a dataset of trajectories, giving it the power to rejecting outliers and identify the likely behaviors of the coupled human-robot system.    

From a safety and interaction perspective, predicting the drivers behavior is incredibly important, as autonomous vehicles are on the precipice of being a part of everyday life.  Here, beyond presenting the algorithm, we focus on two key components of driver behavior: the influence of distraction (i.e. texting while driving) and the impact of intent (i.e. deciding whether or not to change lanes).  
We specifically consider building models of individual driver behavior, but the algorithm presented generalizes across datasets of human-in-the-loop systems, as will be discussed in Section \ref{sec:disc}.  

Frequently, driver monitoring systems that estimate the driver state, are used in Advanced Driver Assistance Systems (typically warning systems) \cite{Doshi2011a,lam2015}.  If we want to take an active and preventative approach by integrating these driver models into control frameworks, a predictive model of the effect on the dynamical system is needed.  Building off the data-driven reachable set concept \cite{driggscampbell2015} and control framework presented in \cite{shia2013}, we aim to create a highly precise and accurate model of human behavior.

The algorithm presented here takes a dataset from a human-in-the-loop system under different conditions (e.g. driver state, environmental conditions, etc.) and outputs empirical reachable sets that have rejected unlikely samples up to some probability threshold.  By selecting observed trajectories of the system, the explicit calculation of the reachable set is estimated by finding the bounds on the dataset, given a mode of behavior.  This can be used to identify mislabeled data or identify the most likely behaviors from the human-robot system.  Due to the flexibility of the modeling method, we can build a more informative and useful reachable set that is usable in a wide variety of scenarios, if represented in the dataset.

The paper is structured as follows. Section \ref{sec:methods} will present our modeling methodology and set prediction algorithm.  To demonstrate the utility of this method, experiments to collect driver data in a realistic human-in-the-loop testbed will be presented in Section \ref{sec:exp_setup}.  The resulting model is be presented and evaluated in Sections \ref{sec:eval} and \ref{sec:apply}.
Finally, we will conclude with a discussion of the results and of the future works.

%% file: sections/modeling_methods.tex
\section{Modeling Methodology}
\label{sec:methods}

As previously described, in this work we aim to develop a framework that when given a dataset of human-robot behaviors, can identify the likely \emph{empirical reachable set}.
First, we state the assumptions, and present the formulation of the problem in terms of finding representative subsets of data.

\subsection{Modeling Assumptions}
In this subsection, we provide the notation and assumptions for the formulation of empirical reachable set to be used as a driver model. 

\noindent \textbf{\emph{Assumption 1: }Existence of Human-in-the-Loop Dataset.} \\
Consider a vehicle with a set of dynamics: 
\begin{equation}
x[k+1] = f(x[k],u[k])
\end{equation}
where $x[k]\in\mathbb{R}^n$ is the state of the vehicle, $u[k]\in U $ is the vehicle inputs where $U \subset \mathbb{R}^m$ is a compact, connected set, $k\in \{ 0,\dots, T \}$ denotes the time step, and $T\in\mathbb{N}$ is the finite time horizon.  
Since we are interested in the human-in-the-loop system, we suppose that the input $u$ comes from the human driver.
The $u$ the human driver uses is affected by the mode, which will be described in Assumption 2.  

For now, we will consider the case where the dynamics $f$ of the vehicle and/or the human control input $u$ is unknown, but we can observe trajectories and recover the observable states.
Suppose we have a dataset $X$ that consists of $N$ sample trajectories of the system over a given time horizon $T$:
\begin{equation}
\begin{array}{cc}
X = \begin{bmatrix} x_1[0] & \dots & x_1[T] \\
\vdots & \vdots & \vdots \\
x_N[0] & \dots & x_N[T] \\       
\end{bmatrix}
\end{array}
\label{eq:X}
\end{equation}
where $x_i[t]\in\mathbb{R}^n$ is a trajectory indexed by $i$.
It is assumed that the initial positions of these trajectories are centered, meaning $x_i[0]=x_0$, $\forall i$.
For notational simplicity, we will denote a sample trajectory as $x_i:=\left[x_i[0] \; \dots \; x_i[T]\right]$.

\noindent \textbf{\emph{Assumption 2: }Existence of Scenario Modes.}  \\
We also assume that we have associated observations of the surrounding vehicles and the environment (e.g. data from radar and road sensors), which we can use to create environmental abstractions.

Suppose given the current sensor information, $e_t$, and the dataset of past observations (or environment abstractions) $E$, we are able to map the current scenario to a past similar scenarios or mode.
\begin{equation}
\theta : e_t \times E \mapsto m
\label{eq:driver_state}
\end{equation}
where $m\in\mathcal{M}$ denotes a driver mode that is one of a finite set of scenario modes that the vehicle could be in.  

This is similar to the hybrid systems formulation where we identify the current mode of operation. 
We can associate this with a set of mental states for the driver (e.g. attentive or distracted), as presented in \cite{krdc2015}, and/or states of the environment.

\noindent \textbf{\emph{Assumption 3: }Existence of Distinct Behavior Modes.}  \\
Given that driver behavior heavily depends on context and that we can identify this mode through $\theta(e_t,E)$, we assume that these modes have associated behaviors and that these behaviors are \emph{unimodal}.    
As was previously mentioned, we are interested in long time horizon trajectory predictions that will encapsulate the uncertainties and the bounds of the potential future states of the vehicle.  

For a particular mode $m \in \mathcal{M}$, we will present an algorithm to calculate some function:
\begin{equation}
\mathcal{A}(X,m) \to \Delta_m(\alpha)
\end{equation}
where we have some function $\mathcal{A}$ that utilizes the dataset $X$ to produce a prediction set as $\Delta_m(\alpha)$.  For a given $\alpha$, the set will encompass the $\alpha$-likely trajectories for mode $m$, as identified by $\theta$. 
The formulation and algorithm to identify this set will be presented in the following section.

\begin{figure*}[!h]
	\centering
	\includegraphics[width=\textwidth]{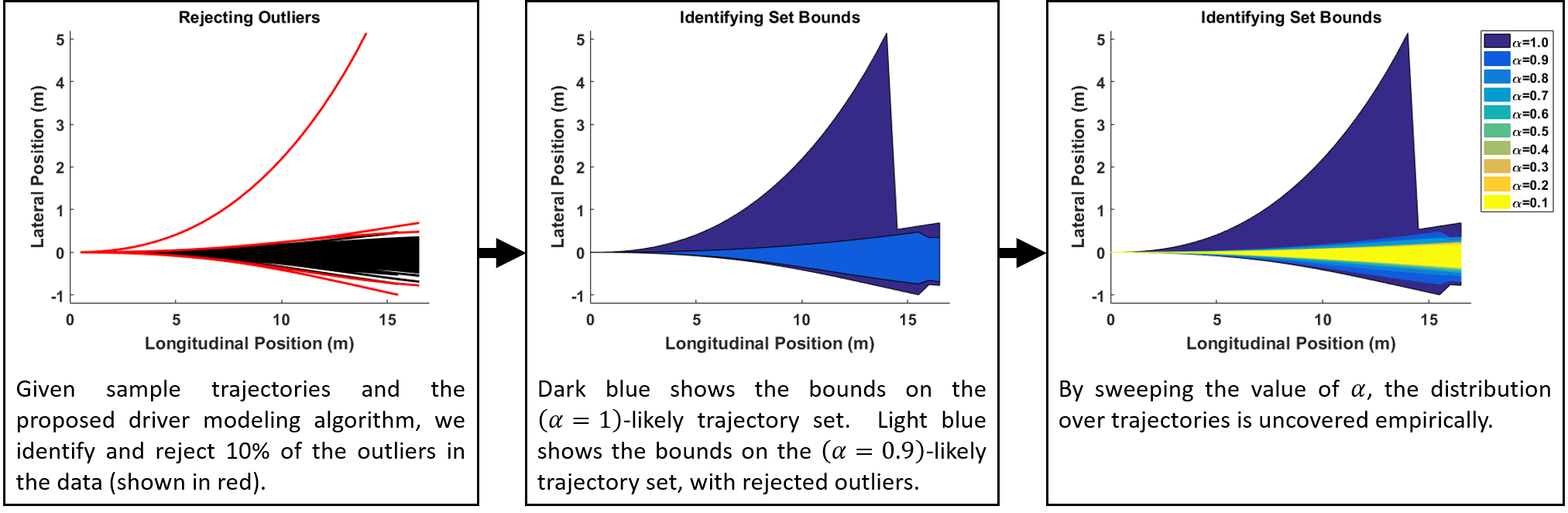}
	\caption{\small \textbf{ Driver Modeling Algorithm.} This flowchart shows how the dataset of trajectories (left) with some outliers (labeled in red) becomes disturbance bounds with some probability threshold (right). For all trajectories, the initial position is centered at (0,0), heading in the positive $x$ direction.  The center image shows the initial set and the new, more precise set with the outliers in red rejected.  The right image shows the full empirical distribution over the dataset.} 	\label{fig:dm_alg}
\end{figure*}

\subsection{Identifying the Empirical Reachable Set}
\label{sec:ERS}

In order to approximate the reachable set and give a reasonable prediction of the system, we present an algorithm (previously denoted $\mathcal{A}$) for deriving the empirical reachable set with outlier rejection to capture the \emph{likely} behavior of the system.

To find a more useful representation of this dataset, we'd like to find the minimum area set that contains the $\alpha$-likely trajectories.
Formally:
\begin{equation}
\begin{array}{ll}
\text{argmin}_{\Delta\subseteq\mathbb{R}^{T}} & \lambda(\Delta)  \\
\text{subject to } & \hat{P}_X(\Delta)\geq \alpha  
\end{array}
\end{equation}
where $\Delta$ is the predicted set, $\lambda(\cdot)$ is the Lebesgue measure that gives the size of the set, and $\hat{P}_X(\Delta)\in\sigma(\mathbb{R}^{T})$ is the empirical probability over the trajectories in dataset $X$:
\begin{equation}
\hat{P}_X(\Delta) = \frac{1}{N}\sum_{i=1}^{N} \mathbb{I}\{x_i\in\Delta\}   
\end{equation}

Since our primary concern is interaction and safety in terms of constraints on the vehicles motion, the trajectories of the high dimensional dynamics are projected into $\mathbb{R}^{n_c}$, where $n_c = 2$, to capture vehicle position.

To make this optimization more concrete, we rephrase the problem as a mixed integer linear program (MILP) that minimizes the area between two bounding hyperplanes that select a subset of the trajectories to meet the probability threshold:
\begin{equation}
\begin{array}{ll}
\AM{\ubar{x},\lbar{x}\in\mathbb{R}^{T}} & \text{area}(\ubar{x},\lbar{x})  \\
\text{subject to }  & b_i(\ubar{x}-x_i)\geq 0 \\
& b_i(\lbar{x}-x_i)\leq 0 \\
& \sum_i b_i \geq N(1-\alpha) \\
\end{array}
\end{equation}
where $b_i$ is the decision variable associated with trajectory $i$.  
This decision variable chooses whether or not this trajectory will be included in the set or not, ensuring that more than $N(1-\alpha)$ trajectories are included.
The area between the bounding hyperplanes $\ubar{x}$ and $\lbar{x}$ is approximated using the Riemann sum approximation. 

These bounds optimally result in being the pointwise minimum and maximum over the subset of the data, determined by the constraints.
Simply put, this algorithm identifies the lines that bound the most precise subset of the trajectories that captures the $\alpha$-likely trajectories behaviors.

However, this is a bilinear constraint and is therefore not easily solvable.
In order to make these constraints linear, we use a cute trick to recast the constraints as linear equations:
\begin{equation}
\begin{array}{cc}
\ubar{x}-x_i \geq (1-b_i)(x_{min}-x_i)   \\
\lbar{x}-x_i \leq (1-b_i)(x_{max}-x_i) 
\end{array}
\end{equation}
where $x_{min}$ is the pointwise minimum and $x_{max}$ is the pointwise maximum of the dataset.
This changes the decision variables to select when the trajectory will be included in the set and when a trivial constraint will be satisfied (i.e. when the upper boundary $\ubar{x}$ is greater than the minimum of the set).

By casting this problem as a mixed integer program, we can efficiently solve for the set that will allow us to choose the trajectories in the data that maximize the precision, given an empirical probability threshold. 
This formalization allows us to capture likely behaviors of the system, and reject outliers from the dataset to derive a more precise and useful trajectory prediction set.

\subsection{An Example}
Suppose we have a dataset consisting of sample trajectories of a human driver lane keeping, which may consist of some outliers, as visualized in the first panel of Figure \ref{fig:dm_alg}.

By simply taking the pointwise bounds of the samples, a conservative estimate of the empirical set is found (as was done in \cite{krdc2015}).
However, we would like to find a representative set and use the algorithm to reject the outliers.
Using the algorithm presented and allowing a rejection of up to $10\%$ of the samples, the optimization program identifies the outliers, as labeled in red sample trajectories in center panel of Figure \ref{fig:dm_alg}.
By sweeping over the probability threshold, the empirical reachable sets with are identified.

We note two key points.
First, an interesting observation can be made by looking at the effect of the precision as we vary probability thresholds, as seen in the right-most figure in Figure \ref{fig:dm_alg}.
A sort of invariant set appears, when the precision no longer changes significantly by throwing out more samples (this will be discussed in more detail shortly).

Second, this data is associated with a specific mode of operation or scenario.
In general, this algorithm assumes a unimodal data distribution, to uncover the the most precise, representative subset of the data.
Our method of overcoming this will be discussed in Section \ref{sec:apply}.
Much like the reachable set analysis utilized by the hybrid systems community, the power of this tool comes from looking at modes which will determine the high level control actions of the human.

%% file: sections/algorithm_evalution.tex
\section{Algorithm Evaluation}
\label{sec:eval}

To demonstrate the functionality of the algorithm in a tangible ground truth, the method is employed on known distributions to make sure these empirical sets are providing useful set with respect to probability thresholds in a reasonable computation time.

\subsection{Distribution Analysis}

To exemplify and validate algorithm performance, baseline results on known distributions were performed for uniform, normal, extreme value, and log-normal distributions.
These were selected to span a range of distributions with varying likelihood of outliers.

\begin{figure*}[!h]
	\centering
	\includegraphics[width=\textwidth]{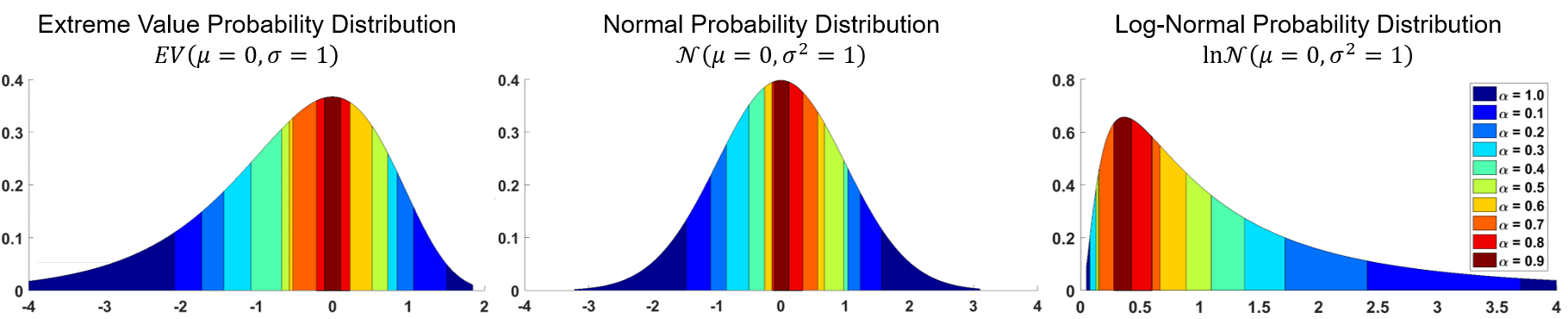}
	\caption{\textbf{Distribution Analysis of ERS Algorithm.}  Sweeping over probability thresholds, we approximate the distribution using data sampled from known distributions.}
	\label{fig:all_sets}
\end{figure*}

To test this, $N$ data-points were drawn at random from each distribution.  
These datasets are passed through the ERS algorithm to identify sets that capture the most precise subset of $\alpha N$ samples of the data.
Samples of the sets overlayed with the probability density functions for the extreme value, normal, and log-normal distributions are visualized in Figure \ref{fig:all_sets}.

\begin{figure}[h!]
	\centering
	\includegraphics[width=\columnwidth]{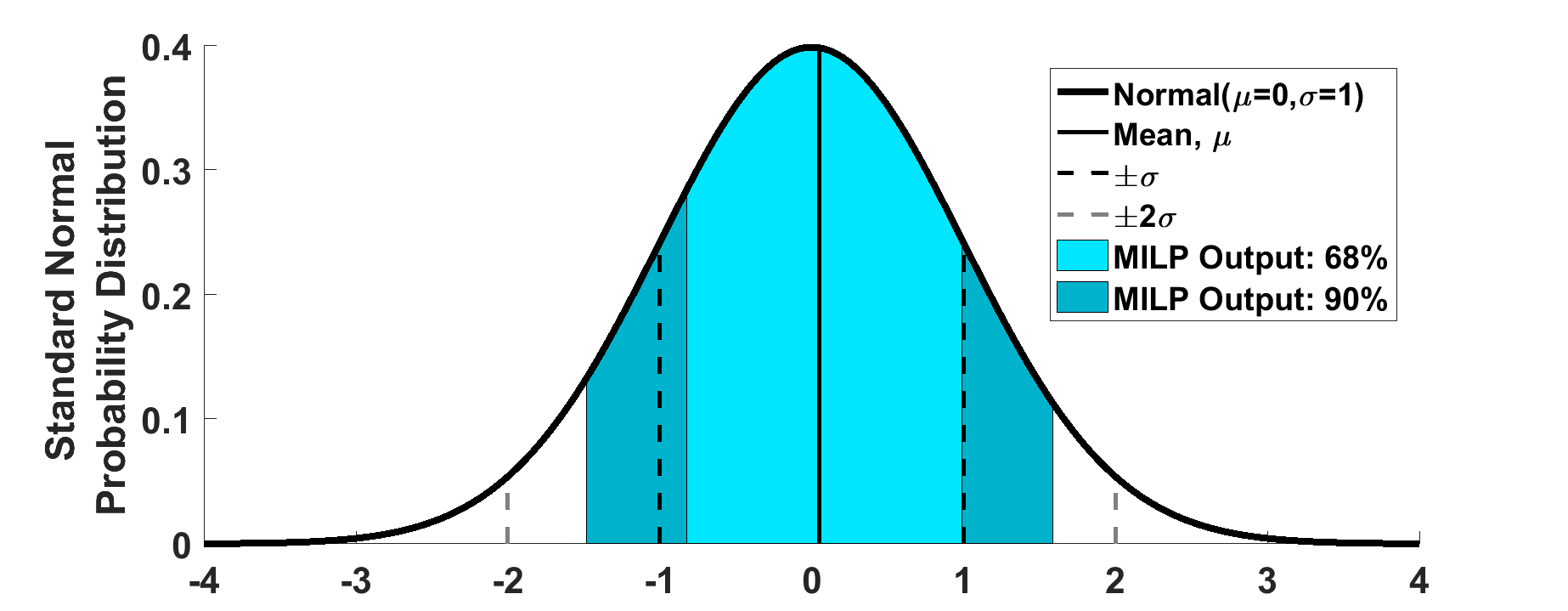}
	\caption{\textbf{Comparing ERS Method to Known Standard Deviation Metrics.}  Given a normal distribution, this plot visualizes and compares our results with the known sets associated with one and two standard deviations.}
	\label{fig:sigmas}
\end{figure}

It can be observed that this method tends to capture the high density regions well and quickly rejects the extreme samples from the sets.  
For a more quantitative sanity check, a normal distribution, the sets found to capture the 1 and 2-$\sigma$ bounds that capture 68\% and 90\% of the data, respectively, in Figure \ref{fig:sigmas}.
We observe that the sets match the tighter bound quite well, but the 2-$\sigma$ bound has some error.
This is to be expected, as the empirical data is more likely to appear near the mean, capturing the typical sets, rather than the distribution itself.

%\begin{wrapfigure}{R}{.6\textwidth}
%\end{wrapfigure}

\begin{figure}[!h]
	\centering
	\includegraphics[width=\columnwidth]{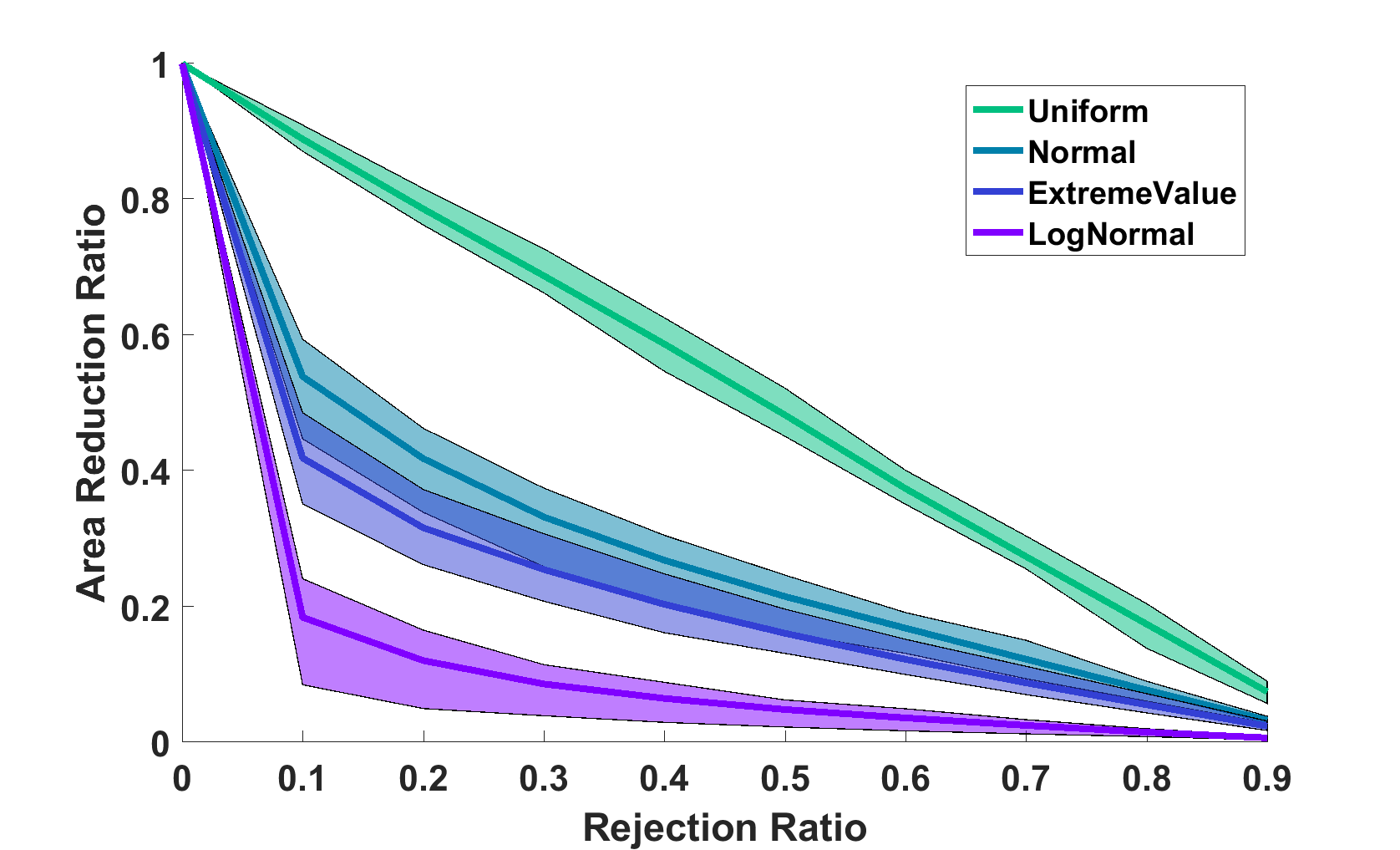}
	\caption{\textbf{Change in Area and Typical Sets.}  This plot shows the change in the size of the set as more and more samples are rejected from the sets.  The lines for each distribution tested show the average area reduction and the maximum and minimum bounds are shown by the shaded regions.}
	\label{fig:area}
\end{figure}

If we consider the size of the sets for each distribution, we see the trend illustrated in Figure \ref{fig:area}.
This plot shows the area reduction ($\delta A = \lambda(\Delta(\alpha))-\lambda(\Delta(\beta))$), where $\beta \in [0 \; 1]$ and $\alpha>\beta$, for each of the distributions.
We note that this value is normalized for plotting convenience and that the $x$-axis shows the \emph{rejection ratio}, which is $1-\alpha$ and ($1-\beta$) in the above notation.

We can see the shape of this curve is dependent on the likelihood of outliers in the data.
For the uniform distribution, we have a linear relationship between the size of the set and number of samples rejected (or the number remaining in the set).
For the Log-Normal distribution, we see that after rejecting the extreme outliers, the size of the set approaches a steady state, where the subset nearly remains the same.
Using these observations, we see that a \emph{typical set} can be identified when we have diminishing returns on the objective function.

%\begin{wrapfigure}{R}{.5\textwidth} 
%\end{wrapfigure}

\subsection{Comparing Efficiency}

As an additional evaluation, the computational efficiency of the MILP formulation of the problem compared to naive approaches for rejecting outliers, without assuming an underlying distribution.
A Leave-$k$ Out method was implemented for comparison:
\begin{equation}
[A^*,i^*] \leftarrow \min\left(\lambda(X|_I)\right)
\end{equation}
where $A^*$ is the minimum area associated at index $i^*$. All combinations of the trajectory indices is given by the $N$-Choose-$N$-$k$ combinatorial function, where $N$ is the total number of samples and $k$ is the number to reject, is denoted $I\in \mathbb{N}^{N_c\times N-k}$.  Each row contains one of the combinations of trajectories to be included in the set.
The minimum function returns the minimum area subset given the areas for all $N_c$ enumerated areas, given by $\lambda(\cdot)$. 

\begin{figure}[!h]
	\begin{center}
		\includegraphics[width=\columnwidth]{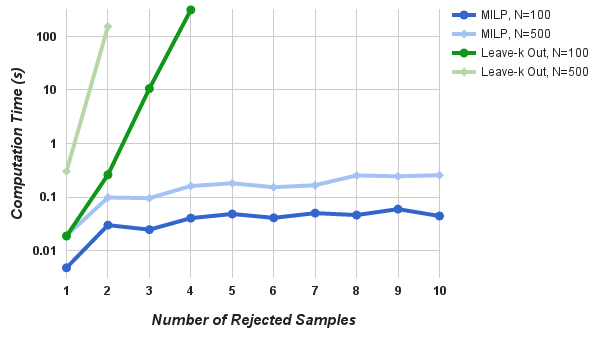}
		\caption{\textbf{Efficiency of the MILP Implementation Compared to Naive Approaches.}  Computation time of rejecting increasing numbers of samples for the two different implementations are shown for $N=100$ and $N=500$.  Our approach is shown in the blue lines and the naive approach is shown in green.}
		\label{fig:compare_timing}
	\end{center}
\end{figure}

The computation time for our formulation and the leave-$k$ out method are shown in Table \ref{tab:timing} and Figure \ref{fig:compare_timing}.

\begin{table}[!h]
	\caption{\textbf{Comparing Computation Time for MILP Implementation and Leave-$k$ Out Approach}.  Results showing computation time in seconds for increasing number of samples to be rejected.}
	\begin{center}
		\begin{tabular}{|l||c|c|c|c|c|c|} %l|}
			\hline
			\multicolumn{7}{|l|}{N=100} \\
			\hline
			\multicolumn{1}{|r|}{k} & 1 & 2 & 3 & 4 & 5  & 10 \\ \hline 
			ERS & 0.005 & 0.030 & 0.025 & 0.041 & 0.048  & 0.44 \\ 
			\hline 
			Naive & 0.019 & 0.261  & 10.572  & 310.917 & --- &	--- \\  \hline \hline
			\multicolumn{7}{|l|}{N=500} \\
			\hline
			\multicolumn{1}{|r|}{k} & 1 & 2 & 3 & 4 & 5  & 10 \\ \hline 
			 ERS & 0.019 & 0.098 & 0.095 & 0.161 & 0.181 & 0.257 \\ 
			\hline 
			Naive  & 0.303 & 152.352 &--- & ---& --- &---	 \\ 
			\hline
		\end{tabular}
	\end{center}
	\label{tab:timing}
\end{table}

We note that the complexity of this optimization problem grows as more trajectories are included in the dataset and generally as more trajectories are rejected.
For the former point, the complexity has not proven itself a serious concern compared to other set-based approaches, due to the assumptions made and the straightforward formulation.
The latter point is intuitive, due ot the fact that region is more dense making the decision variables more difficult to optimally identify.
In some cases (e.g. $N\gg0$, $X$ is quite dense), high values of $\alpha$ are intractable in reasonable time.

To speed up the process of estimating the prediction set over varying values of $\alpha$, we can take advantage of the fact that these sets are \emph{submodular}. 
This means that $\Delta_m(\alpha_p)\subseteq\Delta_m(\alpha_q)$ for all $\alpha_p\leq\alpha_q$.
In practice, this means we can more efficiently calculate the prediction set in dense regions (i.e. high values of $\alpha$, denoted $\bar{\alpha}$), by iteratively computing decreasing values of $\alpha \in [\bar{\alpha} \; 1]$ and reducing the search region to the area from the previous set.
This allows us to chip away at the problem and provides the full approximation of the empirical distribution. 

\begin{figure}[!ht]
	\begin{center}
		\includegraphics[width=.5\textwidth]{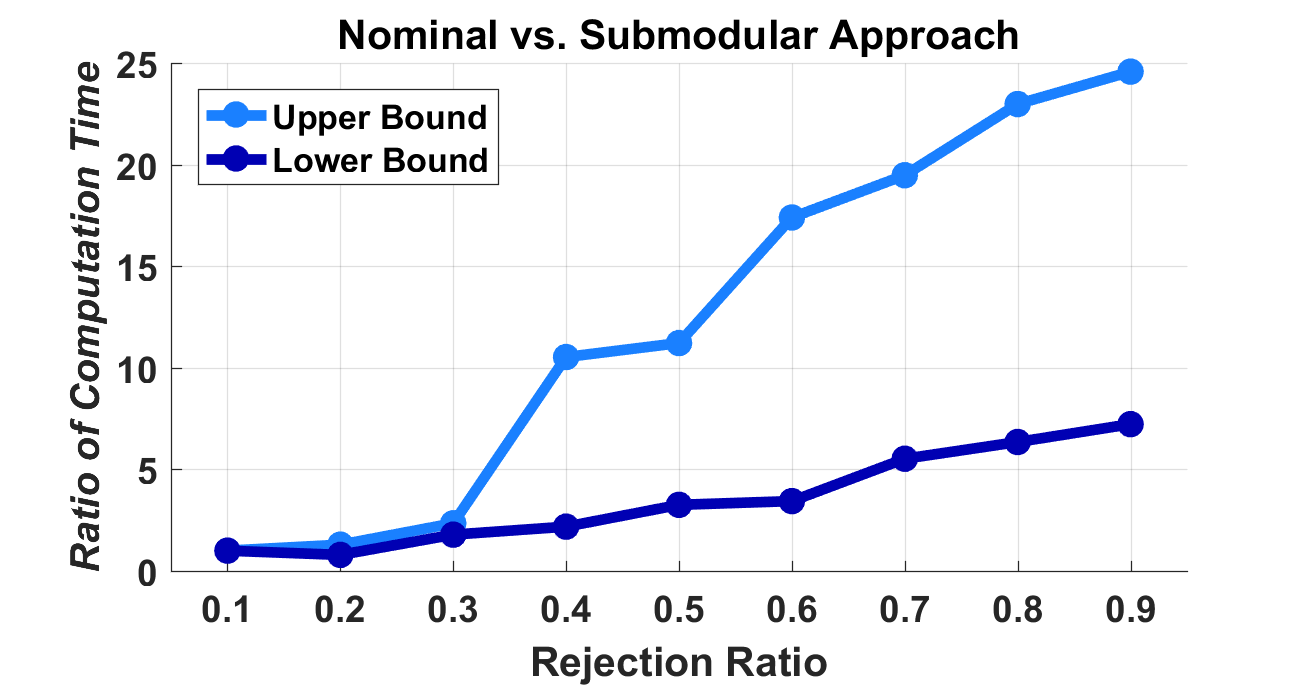}
		\caption{\textbf{Comparing the Submodular Approach to the Baseline Method.}}
		\label{fig:submod_timing}
	\end{center}
\end{figure}

The ratio of the baseline implementation over the submodular approach is shown in Figure \ref{fig:submod_timing}.
As shown, we significantly improve the computation time.
We also note that the plot shows the best case baseline implementation, meaning that we considered the minimum time to compute.
This baseline implementation frequently timed out at 10 minutes, due to the underlying complexity of these dense regions.

Moreover, this integer program can be relaxed to penalize deviations from the typical trajectories instead of requiring strict constraints and the number of trajectories can be weighted by importance or similarity to reduce the number of samples.
Such tricks would improve the computation, but further details and implementations are left as future work. 

\subsection{Model Validation Metrics}
\label{sec:valid}

The following section presents our performance metrics that validate the utility of this model, providing metrics for evaluating the informativeness and robustness of the method.
The performance metrics describe how well we can predict human-in-the-loop behaviors.

We introduce two performance metrics to validate this model with respect to the trajectories of vehicle B:

\noindent {\emph{1) Accuracy Metric:}} Does the actual trajectory lie within the prediction set? \\
\noindent {\emph{2) Precision Metric:}} How informative is this predictive set when compare to a generic set prediction?

In essence, we would like to verify that we are reliably predicting driver behavior, and that we are using a set prediction that is relatively small and informative.

We formalize these metrics in the following equations.
Accuracy is defined as:
\begin{equation}
\begin{array}{ll}
A = \frac{1}{N_T}\sum_{j=1}^{N_T}\mathbb{I}\left\{x_j\in\Delta(\theta_*(e_j,E),\alpha)\right\}
\end{array}
\label{eq:ACC}
\end{equation}
where $N_T$ is the number of samples in the validation set with elements $x_i$, $\Delta(\cdot,\cdot)$ denotes the ERS, where the mode $m$ is determined by $\theta_*(e_j,E)$ for each sample for a given $\alpha$.

Precision is defined as:
\begin{equation}
P = \frac{1}{N_T}\sum_{j=1}^{N_T} 1-\frac{\lambda(\Delta(\theta_*(e_j,E),\alpha))}{\lambda(\mathcal{R}(v_j))}
\end{equation}
where $\mathcal{R}(v_j)$ represents a constant velocity reachable set of the vehicle, given the current velocity $v_j$ and all other variables are as previously described.

This gives us an idea of how informative the set is by assessing how much we are shrinking the set from this generic set.
If the precision metric is 1, the size of the prediction set is 0, meaning that we have precisely predicted the exact trajectory.
If the precision metric is 0, then we are not reducing the size of the set and we can surmise that this prediction is not informative.

%% file: sections/application.tex
\section{ERS on a Lane Changing Example}
\label{sec:apply}

As previously described, we would like to follow the example of the hybrid systems community, we introduce the concept of driver modes to apply this approximation of the reachable set to human modeling.
It is assumed that the way a human controls a system is dependent upon a number of different influences.
For example, a driver will behave differently if they are attentive or distracted, or if they are planning on staying in their lane or executing a lane change.

In our formulation, these modes will come with different partitions in the dataset, as guided by the scenario mode function $\theta$.
This means that by identifying modes of behavior and collecting a sufficient dataset corresponding to that mode, that we can build these predictive sets to represent these different behaviors.
We consider the affects on the predictive sets who's modes change based off of driver intent, as example of how to identify typical sets.

To do this, we will explore different ways of determining modes of intent.
In this formulation, the prediction set $\Delta_m(\alpha)$ and the associated mode $m$ is determined by the driver mode identification function, $\theta(e_t,E)$. 
There are many different methods for determining modes of behaviors, including supervised approaches that predict specific predetermined modes and unsupervised approaches to identify natural groupings within the observed data.

\subsection{Experimental Setup}
\label{sec:exp_setup}
To apply our method for predicting human-in-the-loop behaviors in the context of driving, we must collect our trajectory dataset $X$ and for each method of detecting driver modes.
We collected 1000 sample lane changes from ten subjects.

The resulting dataset consisted of lane changing maneuvers in dynamic environments with up to three vehicles.
For simplicity, we examine a simple scenario of driving in a two-lane, one way road, in a non-urban setting with a varying number of vehicles.
Multiple scenarios were created in which the driver traverses a straight two lane road attempting to maintain a speed between $15$ and $20$ m/s.
Scenarios were generated by creating combinations of the simulation parameters to collected a complete dataset.  
The following parameters were varied: \\
(1) the initial speed and lane location of ego vehicle; \\
(2) the number and location of surrounding vehicles, varied from one to three; and \\
(3) the initial and final speed of each surrounding vehicle.

For example, in some scenarios, the lead vehicle would slow down, forcing the driver to change lanes only if there was room in the next lane. 
Thus, the key here is finding the configurations of the environment states that cross the boundary or safety margin of the human and allows us to identify their likely action between staying in the lane (i.e. braking) or changing lanes to maintain her desired speed.
We note that some scenarios did not require a lane change (e.g. the relative speed of the lead vehicle was initialized such that the driver never felt the need to overtake them), while other scenarios which heavy traffic caused multiple lane changes, but varied depending on the driver's behaviors in the simulation. 

To take a supervised approach, we actively label the data using the driver's turning signal (blinker) to determine when the transition between the \emph{lane keeping} mode and \emph{lane changing} mode.
By using the driver input to label the signal, the driver's thought process is captured and arbitrary heuristics are avoided.
This transition generally occurred one to two sections prior to exiting the lane.
This means that learning these transitions inherently capture a predictive model, meaning that the lane change will be predicted prior to the maneuver actually occurring.

\begin{figure}[h!]
	\centering
	\includegraphics[width=.9\columnwidth]{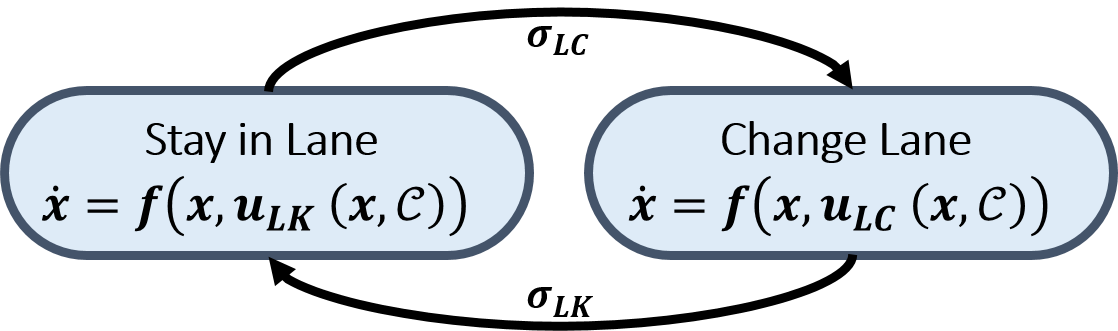}
	\caption{\textbf{Discrete states of Driving Example.} Illustration of discrete modes in our hybrid model of driver intent, where we model the transitions as discrete inputs, $\sigma_*$.}
	\label{fig:modes}
\end{figure}

\subsection{Mode Identification}

\begin{figure}[!b]
	\centering
	\includegraphics[width=.8\columnwidth]{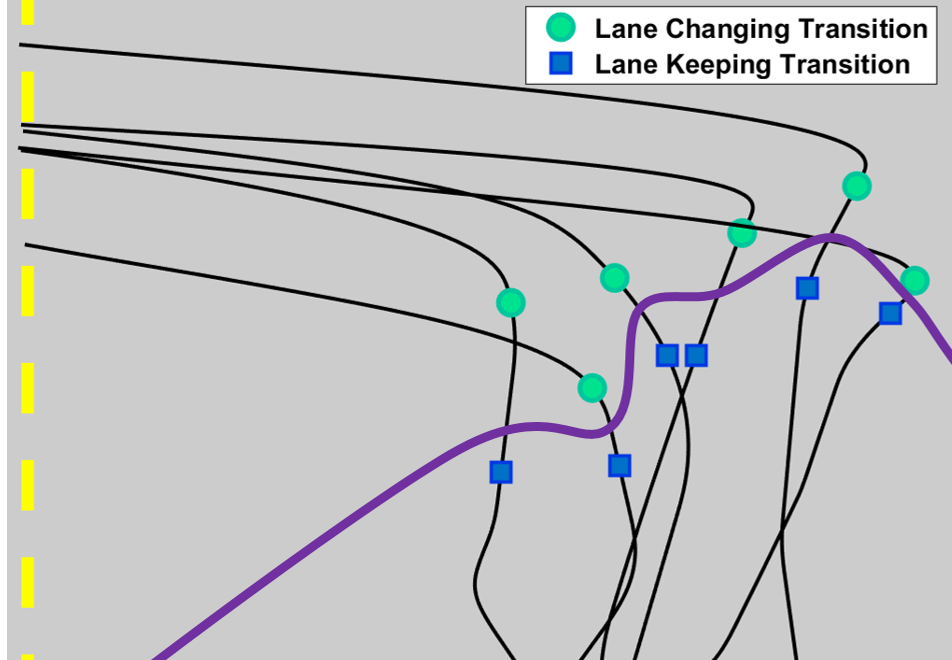}
	\caption{\textbf{Learning Separating Hyperplane between Modes of Intent.}  Sample trajectories of lane changes are shown as the black paths, with the transition points labeled: lane keeping ends at the square points and the beginning of lane changing are shown as circles.  The learned separating hyperplane is shown in purple.}
	\label{fig:learning}
\end{figure}

\begin{figure*}[!h]
	\begin{center}
		\includegraphics[width=\textwidth]{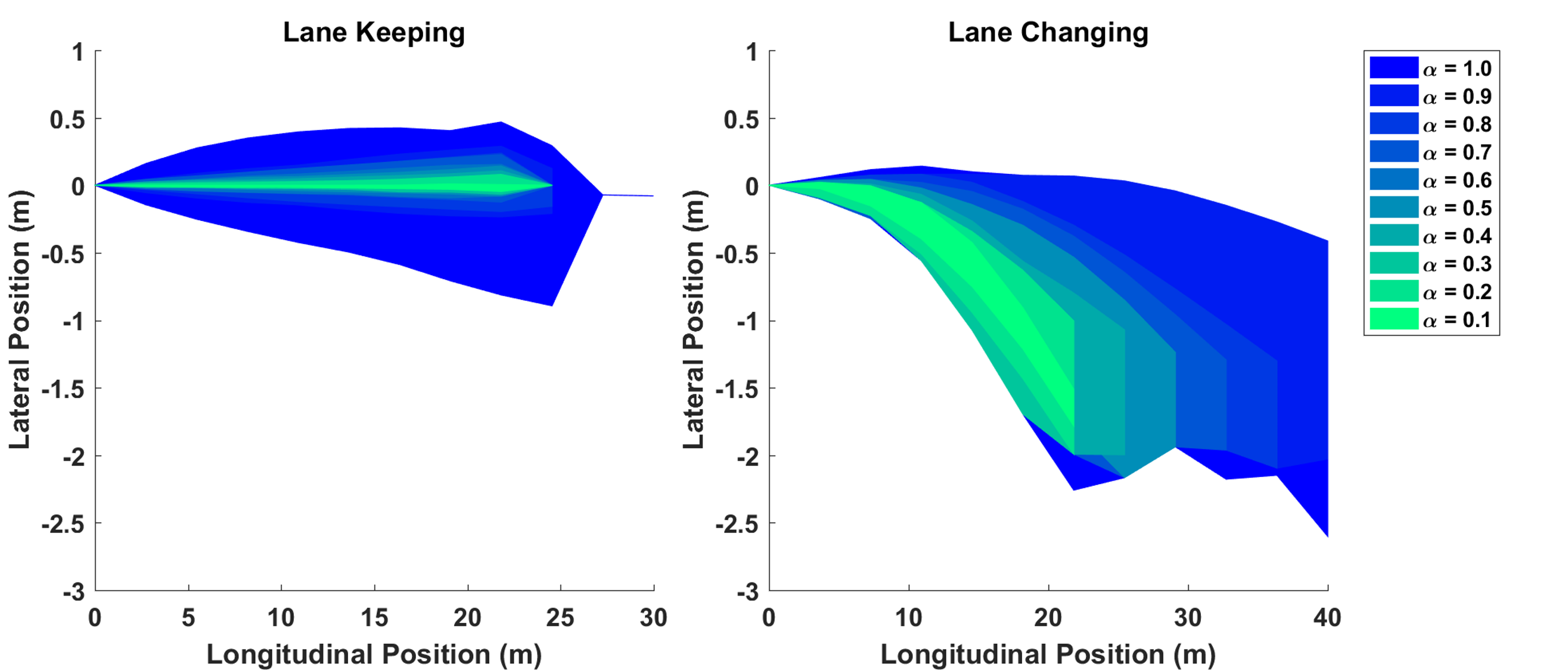}
		\caption{\textbf{ERS Output for Lane Changing Modes.} The empirical set predictions are provided for varying $\alpha$ values.}
		\label{fig:pred_modes}
	\end{center}
\end{figure*}

Detecting lane changes from a dataset has traditionally been done by determining when a lane change occurs by some heuristic (e.g. when the heading angle passes a particular threshold or when the vehicle exits the lane).  
These models look at the data leading up to this point in order to predict that a lane change will occur in the next few seconds \cite{Doshi2011a,Kuge2000}.
This, however, does not capture the decision making process of the human, or capture the idea that these decisions occur as a function of the environment, not just time.
In the proposed detection method, we choose to only rely on the state of the environment, not a predetermined time horizon, meaning that the resulting model allows the prediction time horizon to change.
Further, this approach captures typical human interactions.
While drivers often rely on turning indicators or blinkers to convey our intent to surrounding vehicles, humans can estimate intent without these visual cues, just by observing the motion of nearby vehicles \cite{krdc2016}.

To do this, given some sample data from sensors, $e_k$, we wish to uncover the driver mode identification function $\theta(e_k,E)$ given previously observed data in $E$. 
Given that we are interested in simple lane changing maneuvers (finer analysis of lane changes will be covered in the following chapter), we wish to identify when the driver transitions from \emph{lane keeping} to \emph{lane changing}.
This mapping is uncovered using classification techniques. Many different tools were examined, and many existing approaches demonstrated similar results.  For simplicity, we will generally discuss classification techniques, focusing on support vector machines which aim to uncover the separating hyperplane between the behavior modes \cite{krdc2015,chang2011}.

Classifying data is a common task in machine learning. Suppose some given data points each belong to one of two classes, and the goal is to decide which class a new data point will be in. 
While there are many hyperplanes that might separate the data, support vector machines aim to select the one with the largest margin between data samples.  
This implies that the distance from the boundary to the nearest data point on each side is maximized. 
This resulting separation boundary is known as the maximum-margin hyperplane and provides optimal stability for the classifier \cite{hsu2003practical}.

For lane changing, we wish to uncover the transitions as shown in Figure \ref{fig:modes}.  Using the labeled data previously described, these transition points are learned, and the decision making processes is approximated by these boundaries.
Additionally, each mode will represent a subset of the data that is associated with each mode.  
We will denote this as $X_m = X(I(m))$, where $I(m)$ subset of the dataset that has been associated with mode $m$.

\subsection{ERS Results}
Using this collected data and the intent detection function, we apply the modeling methodology and run the optimization program to identify the probability distribution over trajectories.
The output of this framework are shown in Figure \ref{fig:pred_modes}.

Using the performance metrics defined in the previous section, we validate this method as a predictive model.
The accuracy/precision trade-off curve is shown in Figure \ref{fig:ERS-aVp} in addition to the results provided in Table \ref{tab:ERS-AP}.
Since these sets are determined by a probability threshold (i.e. sweeping over values of alpha), we hope to exceed a baseline performance of a uniform rejection, which is also provided in the figure.
Additionally, to visualize accuracy in a more fluid fashion, the total cumulative error is computed as the total distance between the trajectory and the set prediction, when not on the interior of the set.

\begin{table*}[!h]
	\caption{\textbf{ERS Lane Changing Results}.  Performance metrics and cumulative errors (m) are provided for the two modes presented for lane changes.}
	\centering
	\begin{tabular}{|r||c|c|c|c|c|c|c|c|c|c|} %l|}
		\hline
		\multicolumn{11}{|l|}{\textbf{Lane Keeping Mode}}  \\ \hline \hline
		$\alpha$ & 1.00& 0.90 & 0.80& 0.70& 0.60& 0.50& 0.40& 0.30& 0.20& 0.10          \\ \hline 
		\emph{Accuracy} & 0.98 & 	0.72 & 0.63 & 	0.53 & 0.43 & 0.31 & 0.21 & 0.15 & 0.10 & 0.05 \\ \hline
		\emph{Precision} & 0.64	& 0.72 & 0.79 & 0.84 & 0.88 & 0.91 & 0.93 & 0.95 & 0.98 & 1.00 \\ \hline
		\emph{Cumulative Error} & 0.14 & 0.17& 0.18& 0.21& 0.26& 0.33 & 0.40& 0.47& 	0.57     & 1.48	   \\ 
		\hline\hline
		\multicolumn{11}{|l|}{\textbf{Lane Changing Mode}}  \\ \hline \hline
		$\alpha$ & 1.00& 0.90 & 0.80& 0.70& 0.60& 0.50& 0.40& 0.30& 0.20& 0.10          \\ \hline 
		\emph{Accuracy} & 0.84 & 	0.68 & 	0.66 & 	0.66 & 	0.59 & 	0.48 & 	0.38 & 	0.27 & 	0.24 & 	0.05 \\ \hline
		\emph{Precision} & 0.11 & 	0.43 & 	0.51 & 	0.62 & 	0.66 & 	0.75 & 	0.80 & 	0.83 & 	0.92 & 1.00 \\ \hline
		\emph{Cumulative Error} & 0.99  & 1.49 & 1.67 & 1.69 & 1.76 & 1.90 &  2.09 & 2.14 & 2.83& 2.98 \\ 
		\hline
	\end{tabular}
	\label{tab:ERS-AP}
\end{table*}

Thus, we have produced sets that encapsulate the empirical behaviors of drivers in these modes with reasonable accuracy.
For further nonparametric distribution analysis, we consider the area reduction by rejection ratio as previously shown. 
As expected, we see the same trend as the evaluated heavy tailed distributions.
It can be observed that after rejecting 20\% of the data, we see a nearly linear reduction in the size of the set with each increased step in $\alpha$.
This implies that after this point in the graph, we are no longer rejecting outliers, but eliminating boundary trajectories that have little influence on the precision of the set.
This also matches the templates or footprints from known distributions with outliers, or heavy tails, implying that many driver behaviors are likely members or such distributions.

%% file: sections/discussion.tex
\section{Discussion}
\label{sec:disc}

In summary, this paper presents a novel modeling method that aims to capture driver behavior in a manner that balances informativeness and robustness.  
By using this modeling methodology, we capture the uncertainty in driving behaviors and in a flexible and modular method. 
This not only gives us insight to driving behaviors, but this framework can also be applied to semi-autonomous frameworks, as was shown in \cite{shia2013}.
Future works aim to extend this to interactive scenarios where predictive modeling is required for cooperative control and collaboration.

We note that this method has a few drawbacks.
First, this is highly dependent upon the mode identification and scene classification, meaning that if the dataset is partitioned improperly, you'll likely get uninformative results.
Second, given the cost function currently used (minimizing precision), the algorithm primarily penalizes lateral variation, meaning that it will eliminate trajectories with extreme steering first.
This also means that it favors trajectories that slow down (minimizing the longitudinal distance traveled), which is where most of the errors lay.
More consideration in cost functions is left as future work

Further, there are many other extensions that might improve this work, including taking into account multiple or competing objectives and driver modes.  
Similarly, taking into account bimodal behaviors and mixing intent is a promising avenue for expanding this to more general applications.

\newpage

\begin{figure}[t!]
	\centering
	\includegraphics[width=\columnwidth]{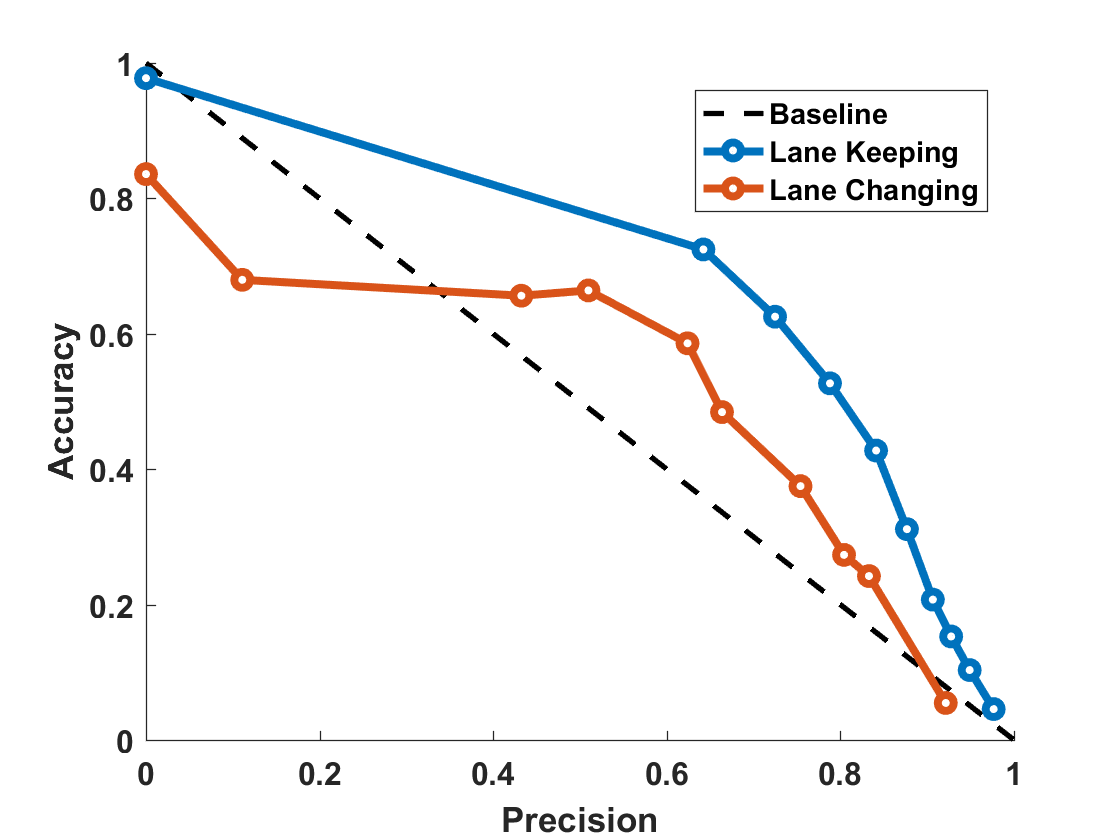}
	\caption{\textbf{Accuracy/Precision Trade-off Curve for Lane Change Example.}  This plot shows the trade-off between accuracy and precision.  The expected uniform baseline performance is shown by the dashed line.}
	\label{fig:ERS-aVp}
\end{figure}
\begin{figure}[t!]
	\centering
	\includegraphics[width=\columnwidth]{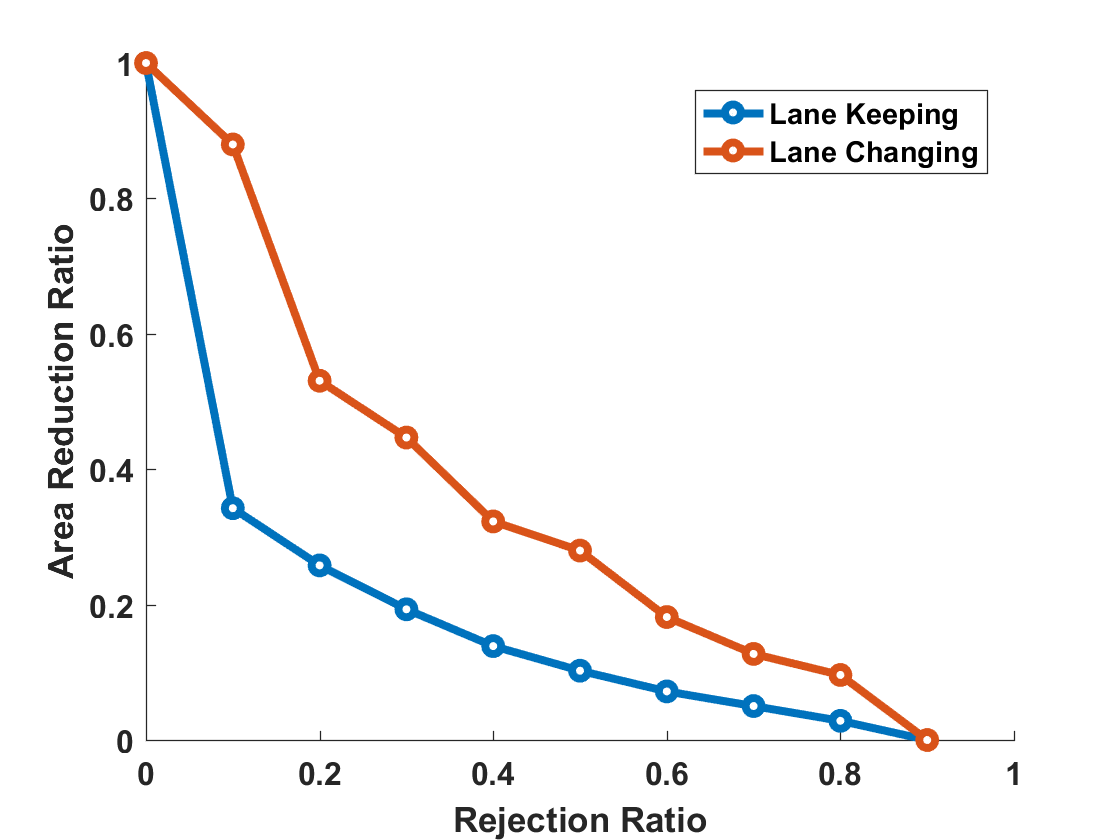}
	\caption{\textbf{Area Reduction Ratio for Lane Change.}  The amount the size of the set decreases by rejection ratio is shown for both modes.  The trend is similar to those of the evaluated heavy tailed distributions.}
	\label{fig:LC-ar}
\end{figure}

%% file: main.bbl
\begin{thebibliography}{10}
\providecommand{\url}[1]{#1}
\csname url@samestyle\endcsname
\providecommand{\newblock}{\relax}
\providecommand{\bibinfo}[2]{#2}
\providecommand{\BIBentrySTDinterwordspacing}{\spaceskip=0pt\relax}
\providecommand{\BIBentryALTinterwordstretchfactor}{4}
\providecommand{\BIBentryALTinterwordspacing}{\spaceskip=\fontdimen2\font plus
\BIBentryALTinterwordstretchfactor\fontdimen3\font minus
  \fontdimen4\font\relax}
\providecommand{\BIBforeignlanguage}[2]{{%
\expandafter\ifx\csname l@#1\endcsname\relax
\typeout{** WARNING: IEEEtran.bst: No hyphenation pattern has been}%
\typeout{** loaded for the language `#1'. Using the pattern for}%
\typeout{** the default language instead.}%
\else
\language=\csname l@#1\endcsname
\fi
#2}}
\providecommand{\BIBdecl}{\relax}
\BIBdecl

\bibitem{baxter2014}
``Rethink robotics: Safety and compliance,''
  \url{http://www.rethinkrobotics.com/safety-compliance/}, 2014, accessed:
  2014-09-27.

\bibitem{Gillula2012}
J.~H. Gillula and C.~Tomlin, ``Guaranteed safe online learning via
  reachability: tracking a ground target using a quadrotor,'' in \emph{2012
  IEEE International Conference on Robotics and Automation}, May 2012, pp.
  2723--2730.

\bibitem{driggscampbell2015}
K.~{Driggs-Campbell} and R.~Bajcsy, ``Identifying modes of intent from driver
  behaviors in dynamic environments,'' pp. 739--744, Sept 2015.

\bibitem{driggscampbell2017}
K.~{Driggs-Campbell}, V.~Govindarajan, and R.~Bajcsy, ``{Integrating Intuitive
  Driver Models in Autonomous Planning for Interactive Maneuvers},'' \emph{IEEE
  Transactions on Intelligent Transportation Systems, Special Edition:
  Applications and Systems for Collaborative Driving}, {To Appear:} 2017.

\bibitem{vasudevan2012}
R.~Vasudevan, V.~Shia, Y.~Gao, R.~Cervera-Navarro, R.~Bajcsy, and F.~Borrelli,
  ``Safe semi-autonomous control with enhanced driver modeling,'' in
  \emph{American Control Conference}, 2012, pp. 2896--2903.

\bibitem{anderson2010}
S.~J. Anderson, S.~C. Peters, T.~E. Pilutti, and K.~Iagnemma, ``An
  optimal-control-based framework for trajectory planning, threat assessment,
  and semi-autonomous control of passenger vehicles in hazard avoidance
  scenarios,'' \emph{International Journal of Vehicle Autonomous Systems},
  vol.~8, no. 2/3/4, 2010.

\bibitem{houenou2013vehicle}
A.~Houenou, P.~Bonnifait, V.~Cherfaoui, and W.~Yao, ``Vehicle trajectory
  prediction based on motion model and maneuver recognition,'' in
  \emph{Intelligent Robots and Systems (IROS), 2013 IEEE/RSJ International
  Conference on}.\hskip 1em plus 0.5em minus 0.4em\relax IEEE, 2013, pp.
  4363--4369.

\bibitem{werling2012optimal}
M.~Werling, S.~Kammel, J.~Ziegler, and L.~Gr{\"o}ll, ``Optimal trajectories for
  time-critical street scenarios using discretized terminal manifolds,''
  \emph{The International Journal of Robotics Research}, vol.~31, no.~3, pp.
  346--359, 2012.

\bibitem{gindele2010probabilistic}
T.~Gindele, S.~Brechtel, and R.~Dillmann, ``A probabilistic model for
  estimating driver behaviors and vehicle trajectories in traffic
  environments,'' in \emph{Intelligent Transportation Systems (ITSC), 2010 13th
  International IEEE Conference on}.\hskip 1em plus 0.5em minus 0.4em\relax
  IEEE, 2010, pp. 1625--1631.

\bibitem{Wiest2012}
J.~Wiest, M.~Höffken, U.~Kreßel, and K.~Dietmayer, ``Probabilistic trajectory
  prediction with gaussian mixture models,'' in \emph{2012 IEEE Intelligent
  Vehicles Symposium}, June 2012, pp. 141--146.

\bibitem{krdc2013}
K.~Campbell, V.~Shia, R.~Vasudevan, F.~Borrelli, and R.~Bajcsy,
  \emph{Probabilistic driver modeling to characterize human behavior for
  semiautonomous framework}.\hskip 1em plus 0.5em minus 0.4em\relax Korea
  University, 2013.

\bibitem{abbeel2004}
P.~Abbeel and A.~Y. Ng, ``Apprenticeship learning via inverse reinforcement
  learning,'' \emph{In Proceedings of ICML}, 2004.

\bibitem{prajna2007framework}
S.~Prajna, A.~Jadbabaie, and G.~J. Pappas, ``A framework for worst-case and
  stochastic safety verification using barrier certificates,'' \emph{IEEE
  Transactions on Automatic Control}, vol.~52, no.~8, pp. 1415--1428, 2007.

\bibitem{shia2013}
V.~Shia, Y.~Gao, R.~Vasudevan, K.~Campbell, T.~Lin, F.~Borrelli, and R.~Bajcsy,
  ``Semiautonomous vehicular control using driver modeling,'' \emph{IEEE
  Transactions on Intelligent Transportation Systems}, vol.~PP, no.~99, pp.
  1--14, 2014.

\bibitem{gillula2013}
J.~H. Gillula and C.~J. Tomlin, ``Reducing conservativeness in safety
  guarantees by learning disturbances online: iterated guaranteed safe online
  learning,'' \emph{Robotics: Science and Systems VIII}, p.~81, 2013.

\bibitem{akametalu2014}
A.~K. Akametalu, J.~F. Fisac, J.~H. Gillula, S.~Kaynama, M.~N. Zeilinger, and
  C.~J. Tomlin, ``Reachability-based safe learning with gaussian processes,''
  in \emph{Decision and Control (CDC), 2014 IEEE 53rd Annual Conference
  on}.\hskip 1em plus 0.5em minus 0.4em\relax IEEE, 2014, pp. 1424--1431.

\bibitem{coelingh2007collision}
E.~Coelingh, L.~Jakobsson, H.~Lind, and M.~Lindman, ``{Collision Warning With
  Auto Brake - A Real-Life Safety Perspective},'' apr 2007.

\bibitem{Doshi2011a}
A.~Doshi, B.~T. Morris, and M.~M. Trivedi, ``{On-Road Prediction of Driver's
  Intent with Multimodal Sensory Cues},'' \emph{IEEE Pervasive Computing},
  vol.~10, no.~3, pp. 22 -- 34, Sept. 2011.

\bibitem{lam2015}
C.~P. Lam, A.~Y. Yang, K.~Driggs-Campbell, R.~Bajcsy, and S.~S. Sastry,
  ``Improving human-in-the-loop decision making in multi-mode driver assistance
  systems using hidden mode stochastic hybrid systems,'' in \emph{2015 IEEE/RSJ
  International Conference on Intelligent Robots and Systems (IROS)}, Sept
  2015, pp. 5776--5783.

\bibitem{krdc2015}
K.~{Driggs-Campbell}, V.~Shia, and R.~Bajcsy, ``Improved driver modeling for
  human-in-the-loop control,'' in \emph{2015 IEEE International Conference on
  Robotics and Automation}, May 2015.

\bibitem{Kuge2000}
N.~Kuge, T.~Yamamura, and O.~Shimoyama, ``A driver behavior recognition method
  based on a driver model framework,'' \emph{SAE Technical Paper 2000-01-0349},
  2000.

\bibitem{krdc2016}
K.~{Driggs-Campbell}, R.~Dong, S.~S. Sastry, and R.~Bajcsy, ``Robust,
  informative human in the loop predictions via empirical reachable sets,'' in
  \emph{In Submission}, January 2017.

\bibitem{chang2011}
C.-C. Chang and C.-J. Lin, ``{LIBSVM}: A library for support vector machines,''
  \emph{ACM Transactions on Intelligent Systems and Technology}, vol.~2, pp.
  27:1--27:27, 2011, software available at
  \url{http://www.csie.ntu.edu.tw/~cjlin/libsvm}.

\bibitem{hsu2003practical}
C.-W. Hsu, C.-C. Chang, C.-J. Lin \emph{et~al.}, ``A practical guide to support
  vector classification,'' 2003.

\end{thebibliography}
